


\font\bigbold=cmbx12
\font\eightrm=cmr8
\font\sixrm=cmr6
\font\fiverm=cmr5
\font\eightbf=cmbx8
\font\sixbf=cmbx6
\font\fivebf=cmbx5
\font\eighti=cmmi8  \skewchar\eighti='177
\font\sixi=cmmi6    \skewchar\sixi='177
\font\fivei=cmmi5
\font\eightsy=cmsy8 \skewchar\eightsy='60
\font\sixsy=cmsy6   \skewchar\sixsy='60
\font\fivesy=cmsy5
\font\eightit=cmti8
\font\eightsl=cmsl8
\font\eighttt=cmtt8
\font\tenfrak=eufm10
\font\sevenfrak=eufm7
\font\fivefrak=eufm5
\font\tenbb=msbm10
\font\sevenbb=msbm7
\font\fivebb=msbm5
\font\tensmc=cmcsc10


\newfam\bbfam
\textfont\bbfam=\tenbb
\scriptfont\bbfam=\sevenbb
\scriptscriptfont\bbfam=\fivebb
\def\Bbb{\fam\bbfam}

\newfam\frakfam
\textfont\frakfam=\tenfrak
\scriptfont\frakfam=\sevenfrak
\scriptscriptfont\frakfam=\fivefrak


\def\eightpoint{%
\textfont0=\eightrm   \scriptfont0=\sixrm
\scriptscriptfont0=\fiverm  \def\rm{\fam0\eightrm}%
\textfont1=\eighti   \scriptfont1=\sixi
\scriptscriptfont1=\fivei  \def\oldstyle{\fam1\eighti}%
\textfont2=\eightsy   \scriptfont2=\sixsy
\scriptscriptfont2=\fivesy
\textfont\itfam=\eightit  \def\it{\fam\itfam\eightit}%
\textfont\slfam=\eightsl  \def\sl{\fam\slfam\eightsl}%
\textfont\ttfam=\eighttt  \def\tt{\fam\ttfam\eighttt}%
\textfont\bffam=\eightbf   \scriptfont\bffam=\sixbf
\scriptscriptfont\bffam=\fivebf  \def\bf{\fam\bffam\eightbf}%
\abovedisplayskip=9pt plus 2pt minus 6pt
\belowdisplayskip=\abovedisplayskip
\abovedisplayshortskip=0pt plus 2pt
\belowdisplayshortskip=5pt plus2pt minus 3pt
\smallskipamount=2pt plus 1pt minus 1pt
\medskipamount=4pt plus 2pt minus 2pt
\bigskipamount=9pt plus4pt minus 4pt
\setbox\strutbox=\hbox{\vrule height 7pt depth 2pt width 0pt}%
\normalbaselineskip=9pt \normalbaselines
\rm}


\def\pagewidth#1{\hsize= #1}
\def\pageheight#1{\vsize= #1}
\def\hcorrection#1{\advance\hoffset by #1}
\def\vcorrection#1{\advance\voffset by #1}

\newcount\notenumber  \notenumber=1              
\newif\iftitlepage   \titlepagetrue              
\newtoks\titlepagefoot     \titlepagefoot={\hfil}
\newtoks\otherpagesfoot    \otherpagesfoot={\hfil\tenrm\folio\hfil}
\footline={\iftitlepage\the\titlepagefoot\global\titlepagefalse
           \else\the\otherpagesfoot\fi}

\def\abstract#1{{\parindent=30pt\narrower\noindent\eightpoint\openup
2pt #1\par}}
\def\smc{\tensmc}


\def\note#1{\unskip\footnote{$^{\the\notenumber}$}
{\eightpoint\openup 1pt
#1}\global\advance\notenumber by 1}

\def\frac#1#2{{#1\over#2}}
\def\dfrac#1#2{{\displaystyle{#1\over#2}}}
\def\tfrac#1#2{{\textstyle{#1\over#2}}}
\def\({\left(}
\def\){\right)}
\def\<{\langle}
\def\>{\rangle}
\def\2pd#1#2#3{\frac{\partial^2#1}{\partial#2\partial#3}}

\def\sqr#1#2{{\vcenter{\vbox{\hrule height.#2pt
        \hbox{\vrule width.#2pt height#1pt \kern#1pt
           \vrule width.#2pt}
        \hrule height.#2pt}}}}

\def\ni{\noindent}
\def\ref #1{$^{[#1]}$}
\def\slash{\!\!\!\!/}
\def\lqq{\lq\lq}
\def\rqq{\rq\rq}

\def\d{\delta}

\def\phys{{\hbox{\sevenrm phys}}}

\def\L{{\cal L}}

\def\R{{\Bbb R}}

\def\psiphys{\psi_\phys}


\pageheight{24cm}
\pagewidth{15.5cm}
\hcorrection{-2.5mm}
\magnification \magstep1
\baselineskip=19pt
\parskip=5pt plus 1pt minus 1pt
%
%
\rightline {MZ-TH/93-03}
\rightline {DIAS-STP-93-04}
\vskip 40pt
\centerline{\bigbold ON QUARK CONFINEMENT}
\vskip 30pt
\centerline{\smc Martin Lavelle{\hbox {$^*$}}{\note{e-mail:
lavelle@vipmza.physik.uni-mainz.de}}
and  David McMullan{\hbox {$^{\dag}$}}{\note{e-mail:
mcmullan@stp.dias.ie}}}
\vskip 15pt
{\baselineskip 12pt \centerline{\null$^*$Institut f\"ur Physik}
\centerline{Johannes Gutenberg-Universit\"at}
\centerline{Staudingerweg 7, Postfach 3980}
\centerline{W-6500 Mainz, F.R.\thinspace Germany}
\vskip 12pt
\centerline{\null$^{\dag}$Dublin Institute for Advanced Studies}
\centerline{School of Theoretical Physics}
\centerline{10 Burlington Road}
\centerline{Dublin 4}
\centerline{Ireland}
}
\vskip 7truemm
\vskip 40pt
{\baselineskip=13pt\parindent=0.58in\narrower\ni{\bf Abstract}\quad
A sufficient condition for the
confinement of quarks is presented. Quarks are shown to be unobservable.
Colour singlets are however, observables. The results of deep inelastic
scattering are discussed. We argue that QCD does not exhibit a
deconfining transition.
\bigskip\bigskip
\ni{\bf PACS No.:}\quad 12.38.Aw\quad 13.60.Hb
\par}
\bigskip
\centerline{Submitted to Physical Review Letters}

\vfill\eject
\noindent Quantum Chromodynamics (QCD) is the theory of strong
interactions\ref{1}. Its success is based on perturbation theory. The
content of the theory is a non-abelian, $SU(3)$, interaction of quarks
and gluons. Evidence for these particles comes from deep inelastic
scattering. The outstanding problem in QCD is that these particles have not
been observed experimentally. This has led to the confinement hypothesis
that only colour singlet objects are observed in nature. In this letter we
will prove that this is indeed the case.

The QCD Lagrangian is
$$
\L=-\tfrac14 F^2 +  \bar\psi(i D\slash-m)\psi\,,
\eqno (1)
$$
where $F$ is the field strength constructed out of the non-abelian
potentials $A$, $D$ is the covariant derivative and
$\psi$ is a fermionic field. This Lagrangian exhibits the
following gauge invariance
$$
\eqalign{
A(x)\to A^g(x) & = g^{-1}(x) A(x) g(x)+ g^{-1}(x) d g(x) \cr
\psi(x)\to \psi^g(x) & = g^{-1}(x) \psi(x)\,.
}
\eqno (2)
$$
Observables must be gauge invariant. From Eq.\thinspace2 we see that the
fermionic fields, $\psi$ and $\bar\psi$, are not observables and thus
cannot be identified with observable quarks. A similar problem for the
electron arises in QED and has been solved by Dirac\ref{2} as we now
explain (see also Ref.\thinspace3).

The physical electron field is given by
$$
\psiphys(x)=\exp\(ig\frac{\partial_i A_i(x)}{\nabla^2}\) \psi(x)\,.
\eqno (3)
$$
{}From the abelian version of (2) this may be straightforwardly seen to be
gauge invariant (or, more properly, BRST invariant\ref{3}) up
to rigid (global) transformations.
This physical field is actually the electron: its
propagator is gauge invariant\ref{4}. In contrast to the usual asymptotic
identification of the electron with $\psi$, this electron has an
electromagnetic charge and creates a Coulomb electric field\ref{2}.

An alternative approach to this is as follows. Consider a fermion
attached to a Wilson line
$$
\psi_{_{\scriptstyle\Gamma}}(x)=\exp\(ig\int_{-\infty}^x
A_\mu(z)dz^\mu\)\psi(x)\,,\eqno(4)
$$
where $\Gamma$ is any contour from the point $x$ to $-\infty$.
Although this is, by construction, gauge invariant, it is dependent
on the arbitrary line $\Gamma$. A physical electron {\it cannot\/}
have this property.
Exploiting the gauge invariance of the theory, we set the unphysical
field $A_0$ to zero. The spatial components may be decomposed into
the physical, transverse components, $A_i^T$, and the unphysical,
longitudinal component, $A_i^L=\dfrac{\partial_i\partial_j
A_j}{\nabla^2}$. This means that $\psi_{_{\scriptstyle\Gamma}}$ may be
written as
$$
\psi_{_{\scriptstyle\Gamma}}(x)=N_\Gamma(x) \psiphys(x)\,,
\eqno(5)
$$
where
$$
N_\Gamma(x) =   \exp\(ig\int_{-\infty}^x A_i^T(z)dz^i\)\,.
\eqno(6)
$$
This gauge invariant normalisation factor contains all the contour
dependence and must be removed for the fermion to have any physical meaning.
We thus recover Dirac's physical electron.

A sufficient  condition for the confinement of
quarks would be to show that no contour and gauge independent generalisation
of Dirac's physical electron can be constructed for the quarks. We
will now demonstrate that this is the case.

Working in a Hamiltonian description\ref{5} of QCD, where
the momenta conjugate to
the potential is denoted by $\Pi(x)$, i.e., such that the fundamental
Poisson bracket is $\{A^a_i(x), \Pi^j_b(y) \}= \d^a_b\d_i^j\d(x-y)$, we see
that if such a field exists it may be written as
$$
\psiphys(x)=V(x)\psi(x)\,,
\eqno (7)
$$
where $V(x)$ is a field dependent element of $SU(3)$. This implies that
under a gauge transformation $V$ must transform as
$$
V(x)\to V^g(x)=V(x) g(x)\,.
\eqno (8)
$$
We now assume that for the fundamental fermions of the system
this $V$ may be taken as a function of the gauge fields only. Thus writing
$$
V(x)=\exp\(i v(x)^a T^a  \)\,,
\eqno (9)
$$
where $T^a$ denotes the Gell-Mann matrices, the infinitesimal version
of (8) is
$$
\{v^b(x),G^a(y)\}=\d^{ab}\d(x-y)\,,
\eqno (10)
$$
where $G^a$ is the infinitesimal  generator of gauge transformations
$$
G^a(x)= (D_i\Pi^i)^a(x)+gJ_0(x)\,,
\eqno (11)
$$
and $J_0$ is the current density. If such a $V$
existed, then Eq.\thinspace10 would imply that $v^b(x)$ is a
gauge fixing condition. We now assume that our fields are chosen so
that,
as far as the gauge group is concerned, we can identify the space
time with $\R\times S^3$, where $S^3$ is the spatial
three sphere. As is well known\ref{6} there is  no such
global gauge fixing in QCD (the Gribov ambiguity). Hence there is no
gauge invariant description of a single quark.
Of course there are observables in QCD, these correspond to gauge invariant
combinations of the fundamental fields; an example is $\bar\psi\psi$.

We stress that the above is a {\it sufficient\/} condition for
confinement, and is not a necessary one. Indeed,  abelian theories
(for example, compact U(1) in three dimensions\ref{7}) may also
confine due to dynamical effects.
We now discuss some further consequences of this proof of confinement.
\bigskip
Locally, that is at small scales or high energies, gauge fixings of the
form (10) can be constructed. Thus at such scales quarks will appear to be
physical. Therefore they can be \lq observed\rq\ in deep inelastic
scattering. In such a local description string like singularities could be
expected.  To find the scale of confinement is, however, a hard dynamical
question.

In order to describe QCD at finite temperature and density our assumptions
on the topology of space time must be replaced by $S^1\times S^3$.
This additional complication of the topology will not alter our
arguments, thus we predict that there is no deconfining transition,
although the scales will change.

Another kinematical account of confinement has been proposed by Kugo
and Ojima\ref{8}. The connection between their work and ours is
unclear to us, in particular they make no reference to the role of
the Gribov ambiguity.
\bigskip
\ni{\bf Acknowledgements} MJL thanks the Dublin Institute for Advanced
Studies for their hospitality and the Graduierten Kolleg of Mainz
University for support. DM thanks the University of Mainz for hospitality.
\bigskip
\ni{\bf References}
\item{1)}{T. Muta, \lqq Foundations of Quantum Chromodynamics\rqq, (World
Scientific, Singapore, 1987).}
\item{2)}{P.A.M. Dirac, \lqq Principles of Quantum Mechanics\rqq, (OUP,
Oxford, 1958), page 302.}
\item{3)}{M. Lavelle and D. McMullan, \lqq A New Symmetry for QED\rqq,
Mainz/Dublin preprint, MZ-TH/93-02, DIAS-STP-93-03, submitted
to Physical Review Letters.}
\item{4)}{M. Lavelle and D. McMullan, \lqq On the Physical Propagators of
QED\rqq, Mainz/Dublin preprint, MZ-TH/93-05, DIAS-STP-93-05, submitted
to Physics Letters B.}
\item{5)}{R. Jackiw, in \lqq Current Algebra and Anomalies\rqq, by
S.B.~Treiman et al. (World Scientific, Singapore, 1987).}
\item{6)}{V.N. Gribov, Nucl. Phys. B139 (1978) 1; I. Singer, Commun. Math.
Phys. 60 (1978)~7.}
\item{7)}{A. Polyakov, \lqq Gauge Fields and Strings\rqq, (Harwood,
Chur, 1987). }
\item{8)}{See for example the discussion in N. Nakanishi and I.
Ojima, \lqq Covariant Operator Formalism of Gauge Theories and
Quantum Gravity\rqq, (World Scientific, Singapore, 1990).}
\bye